\documentclass{ptptex}
\usepackage[dvips]{graphicx}

\title{Decay of Color Gauge Fields in Heavy Ion Collisions\\
and Nielsen-Olesen Instability}

\author{Aiichi \textsc{Iwazaki}}

\inst{International Politics and Economics, Nishogakusha University \\
2590 Ohi, Kashiwa, Chiba 277-8585, Japan}

\abst{We analyze the behavior of unstable modes in the glasma produced in high energy heavy ion collisions, 
using a simple model with effective
homogeneous longitudinal color electric and magnetic fields. 
The unstable modes are approximately described as Nielsen-Olesen unstable modes
under the homogeneous longitudinal gauge fields.
We find that the Nielsen-Olesen unstable modes show properties very similar to those of 
the exponentially increasing unstable modes in the glasma recently shown by Romatschke and Venugopalan.
Although initial gauge fields in the glasma are much stronger than the ones in our model,
they decay with the production of 
the Nielsen-Olesen unstable modes. 
We discuss why we can reproduce the features of the glasma by using the effective homogeneous
weak magnetic fields.
Our analysis supports an idea that 
the decay of the gauge fields in the glasma
is caused by
Nielsen-Olesen instability.}

\begin{document}
\maketitle

\section{Introduction}
In high energy collisions of heavy ions in RHIC or LHC, the most important
ingredients for producing quark gluon plasma ( QGP ) are small x gluons in the nuclei~\cite{1,2,3,4,5}.
The small x gluons with also small transverse momenta are sufficiently dense in the nuclei and so
they may be treated as classical fields produced by glassy large x gluons
even after the collisions of the nuclei. It has been shown that 
longitudinal color magnetic and electric fields of the small x gluons are generated 
initially at the collisions. They are classical fields and evolve classically
according to a color glass condensate (CGC) model\cite{cgc}.
These fields, which are called as glasma, are expected to give rise to QGP by their rapid decay\cite{hirano, iwa}.

In an extremely high energy collision, the thickness of nuclei is nearly zero due to the Lorentz contraction.  
Thus, the initial gauge fields have only transverse momentum perpendicular to the collision axis, but
have no longitudinal momentum ( rapidity dependence ). Such classical gauge fields can not
possess any longitudinal momentum in their classical evolution, since equations of motion of the gauge fields
are invariant under the Lorentz boost along the collision axis\cite{2,3,4}.
Thus, the naive application of
the CGC model, e.g. McLerran-Venugopalan model ( MV model )\cite{cgc}, does not give rise to thermalized QGP.
It has recently been shown\cite{ve}, however, that the addition of small
fluctuations with rapidity, e.g. quantum fluctuations\cite{fuku,review} to the initial gauge field
induces exponentially increasing modes with longitudinal momentum. 
The production of the exponentially increasing modes implies that a process toward
thermalization has started; the decay of the gauge field and the isotropization
of momenta. Although the decay of the classical fields have been clarified in the numerical calculations, 
the physical mechanism of the decay is still unclear.

In this paper we show using a simple model of an initial gauge field
that the decay of the glasma is 
caused by Nielsen-Olesen unstable modes\cite{no,savvidy,iwa,itakura}. 
The modes arise under the initial gauge field 
when small fluctuations around the gauge fields are taken into account.
Especially 
we compare the time evolutions of the unstable modes with those of the exponentially increasing
modes of longitudinal pressure shown in the previous calculation\cite{ve}.
We find that the previous main results   
can be reproduced.  

The initial gauge field in our model is spatially homogeneous and much weaker than 
the fields of the glasma\cite{ve}. The glasma are strong ( e.g. color magnetic field $B\sim O(Q_s^2/g)$ ) and
inhomogeneous ( their coherent length being the order of $\sim Q_s^{-1}$; $Q_s$ is a saturation momentum
written as $Q_s=g^2\mu$ in the ref.\cite{ve}. )
The initial gauge field is chosen to reproduce the time evolutions of the 
instabilities observed in the reference\cite{ve}. It can not reproduce
local behaviors of the instabilities in the transverse space since
it is homogeneous in the space.

Nielsen-Olesen instability is associated with color magnetic fields,
not color electric fields. In the paper we are only concerned with the 
instability of the color magnetic fields. The color electric fields 
produced in heavy ion collisions are
unstable against quark pair creations\cite{miklos} so that they may decay 
sufficiently fast. We do not discuss the instability of the color electric fields.

In the next section, we explain the instabilities observed in the previous simulation\cite{ve}.
In the section(3), we review briefly Nielsen-Olesen instability.
In the section(4), we explain the assumption
used in our model, expecially the relevance of the use of homogeneous color magnetic field
instead of inhomogeneous glasma.
In the section(5), we introduce our simple model for analyzing the decay 
of the glasma in $\tau$ and $\eta$ coordinates.
In the section(6), we compare our results with those obtained in the 
previous simulation\cite{ve}. In the final section,
we conclude our results.

\section{Instabilities in initial gauge fields}

In high energy heavy ion collisions, classical color gauge fields
are generated and are pointed to the collision axis ( longitudinal direction ).
They are color electric and magnetic fields.
They arise at $\tau=0$ just when the collisions occur.
( Here we assume collisions in high energy limit where heavy ions are Lorentz contracted to have 
zero width. Thus, the collisions occur at the instance of $\tau=0$; $\tau=\sqrt{x_0^2-x_3^2}$. )
They have sufficiently large energy densities to produce thermalized quark gluon plasma
in the subsequent their decay. The color gauge fields are coherent 
states of small x gluons with transverse momenta less than a saturation momentum, $Q_s$.
These gluons are described by a model of color glass condensate, e.g. MV model.
More explicitly, the color gauge fields are given initially at $\tau=0$ as functions of
color charge density of large x gluons inside
of heavy ions; the color charge
density is determined with a Gaussian distribution in the MV model. 

Such classical gauge fields are uniform in the longitudinal direction.
On the other hand, they are not uniform in the transverse directions;
the scales of their variations in the directions
are typically determined by the saturation momentum $Q_s$. This is because
they are made of the gluons possessing transverse momenta typically
given by $Q_s$. 
This implies that the fields keep their directions ( parallel or anti-parallel to the
collision axis ) inside of
transverse regions whose widths are given typically by $Q_s^{-1}$. But, they change
their sign outside of the region. 
In this way
their directions are never uniform in the
transverse directions, although they are uniform in the longitudinal
direction.

Anyway, such color gauge fields are given initially in the collisions.
After their production they evolve according to the gauge field equations.
Since the gauge fields equations are invariant under Lorentzs boost along the
collision axis denoted by $x_3$ ( $\eta=\log(\frac{x_0+x_3}{x_0-x3}) \to \eta+\mbox{cont.}$ ),
the color gauge fields have no dependence of $\eta$ in their development. That is, the gauge fields 
are still uniform in the longitudinal direction after the collision. 
In such a circumstance, a numerical calculation\cite{5}
has been performed to show that they evolve smoothly in time $\tau$ and
become weak owing to their expansion. 
Any unstable behaviors of the fields were not found in the calculation.
Non-linearity in the gauge field equations seems not to play
dramatici roles in the evolution of the gauge fields. Actually,
it has been shown\cite{itakura} that their time developments
can be reproduced qualitatively in the linear analysis of
the field equations; especially  
the fields become weak with time as $1/\tau$.
( A kind of "Abelian dominance"\cite{abelian} 
seems to hold for large $\tau$ since self-interactions of gauge fields
become ineffective because of smallness of the fields in the large $\tau$. ) 
We use the feature in order to make our
simple analytical model of the glasma decay. 

Subsequently, a numerical simulation has been performed\cite{ve}
with a slightly modified initial condition. That is, much small
gauge fields depending on $\eta$ are added by hand to the original 
initial gauge fields. Such gauge fields may arise because 
the uniformness in $\eta$ of the initial condition is broken 
in real situations: 
Heavy ion collisions occur with finite energies
or the initial conditions derived in the MV model receive
high order quantum corrections. Both of them give rise to
much small corrections depending on $\eta$ to the initial gauge fields without
$\eta$ dependence. 

The simulation has clarified the existence of unstable modes in
the evolution of the gauge fields; Fourier components in $\eta$ of longitudinal pressure
increase exponentially in
$\tau$. This implies that a component of gauge fields added in the simulation
increases exponentially.
The existence of the modes implies that the initial color electric and magnetic fields
uniform in the longitudinal direction are unstable under the small
fluctuations depending on $\eta$.  

We should note that since gauge field fluctuations depending on $\eta$ are sufficiently smaller 
than the initial gauge fields,
they can be treated perturbatively. That is, the fluctuations
evolve under the background initial gauge fields
without self-interactions. Hence, the analysis of their evolution can be
performed in the linear approximation.

Here, we review characteristic properties of the unstable modes found in the 
numerical simulation. We denote the longitudinal pressure as $P_{\eta}$.
When we denote Fourier components
as $P_{\eta}(k_{\eta},\tau)=\int d\eta \,\,P_{\eta}(\eta,\tau)\exp(ik_{\eta}\eta)$ ( $k_{\eta}$ denotes longitudinal momentum ),
$P_{\eta}(k_{\eta},\tau)$ evolves smoothly in $\tau$ 
when initial gauge fields have no dependence of the rapidity $\eta$.
Obviously, $P_{\eta}(k_{\eta},\tau) \propto \delta(k_{\eta})$ in the case. 
Once gauge fields depending on $\eta$ are added to the initial gauge fields,
$P_{\eta}(k_{\eta},\tau)$ shows exponential increase. 
Namely some unstable modes are excited 
when gauge fields fluctuations depending on $\eta$ are added.
$P_{\eta}(k_{\eta},\tau)$ shows the exponential increase just after
a time $\tau(k_{\eta}) >0$ passes. In other words,
$P_{\eta}(k_{\eta},\tau)$ does not increase exponentially until the time $\tau = \tau(k_{\eta})$. The simulation
shows that $\tau(k_{\eta})\propto k_{\eta}$. 
That is, the component $P_{\eta}(\tau,k_{\eta})$
begins to increase exponentially later than the time when components $P_{\eta}(\tau,k'_{\eta})$ 
( $k'_{\eta}<k_{\eta}$ ) increase.
Hence, we can define a maximum momentum $k_{\eta}(\mbox{max})$ at the instance $\tau$
such that  
the component $P_{\eta}(\tau,k_{\eta}(\mbox{max}) )$
begins to increase exponentially at the instance; 
components $P_{\eta}(\tau,k_{\eta}<k_{\eta}(\mbox{max}))$
have already increased exponentially.
The simulation shows that $k_{\eta}(\mbox{max})$ increases
linearly with time $\tau$.

Another interesting property found in the simulation is concerned with longitudinal momentum
distribution at a time $\tau$, namely $P_{\eta}(k_{\eta},\tau)$.
The distribution has a peak at the value of $k_{\eta}(p)$, which is almost independent
of $\tau$. This implies that the unstable modes with the characteristic longitudinal
momentum $k_{\eta}(p)$ are induced dominantly. 
The peak in the distribution, $P_{\eta}(k_{\eta}(p),\tau)$, increases 
such as $P_{\eta}(k_{\eta}(p),\tau)\propto \exp(\mbox{const.}\tau^{1/2})$.

We should mention that the growth rate of the peak is 
much smaller than the typical time scale $Q_s^{-1}$ in the system.
Thus, it takes a too long time for the small gauge field fluctuations
to become comparable order of magnitude of the initial gauge fields.
This is a bad new for the realization of thermalized QGP through the decay of the gauge fields.

We will show that these characteristic properties can be understood
using Nielsen-Olesen instability.
The point in our analysis is that the background initial gauge fields are approximately
described by Abelian gauge fields and the gauge field fluctuations added 
can be treated perturbatively. 
Based on the approximation, we show that 
the unstable modes found in the numerical simulation
are just Nielsen-Olesen unstable modes.    

\section{Nielsen-Olesen Instability}

We review briefly Nielsen-Olesen instability by using SU(2) gauge theory.
The instability means that homogeneous color magnetic field
is unstable in the gauge theory.
In order to explain it, we decompose the gluon's
Lagrangian
with the use of the variables, "electromagnetic field" 
$A_{\mu}=A_{\mu}^3,\,\,\mbox{and} \,\, \mbox{"charged vector field"}\,
\Phi_{\mu}=(A_{\mu}^1+iA_{\mu}^2)/\sqrt{2}$ 
where indices $1\sim 3$ denote color components,

\begin{eqnarray}
\label{L}
L&=&-\frac{1}{4}\vec{F}_{\mu
  \nu}^2=-\frac{1}{4}(\partial_{\mu}A_{\nu}-\partial_{\nu}A_{\mu})^2-
\frac{1}{2}|D_{\mu}\Phi_{\nu}-D_{\nu}\Phi_{\mu}|^2- \nonumber \\
&+&ie(\partial_{\mu}A_{\nu}-\partial_{\nu}A_{\mu})\Phi_{\mu}^{\dagger}\Phi_{\nu}+\frac{g^2}{4}(\Phi_{\mu}\Phi_{\nu}^{\dagger}-
\Phi_{\nu}\Phi_{\mu}^{\dagger})^2
\end{eqnarray}
with $D_{\mu}=\partial_{\mu}+igA_{\mu}$,
where we have omitted a gauge term $D_{\mu}\Phi_{\mu}=0$. 

We can see that
the charged vector fields $\Phi_{\mu}$ couple with electromagnetic field $A_{\mu}$
minimally through the covariant derivative $D_{\mu}$ and non-minimally
through the interaction term, $ig(\partial_{\mu}A_{\nu}-\partial_{\nu}A_{\mu})\Phi_{\mu}^{\dagger}\Phi_{\nu}$.
When a homogeneous color magnetic field $B>0$ described by $A_{\mu}=A_{\mu}^B$ is present,
we analyze the fluctuations $\Phi_{\mu}$ under the color magnetic field. To do so,
we solve the following equation of $\Phi$ under the background field $B$,

\begin{equation}
(\partial_t^2-\vec{D}^2\mp 2gB)\phi_{\pm}=0
\end{equation}
with $A_j^B=(-Bx_2,Bx_1,0)/2$,
where $\phi_{\pm}=(\Phi_1\pm i\Phi_2)/\sqrt{2}$ and $\vec{D}=\vec{\partial}+ig\vec{A}^B$. 
Index $\pm$ denotes a spin component parallel ($+$) and anti-parallel ($-$) to $\vec{B}=(0,0,B)$. 
We have assumed the magnetic field pointed into $x_3$ direction.
In the equation, higher order interactions have been neglected.
The energy spectra of the fields $\phi_{\pm}$ are easily obtained.
The energy $\omega$ of
the charged vector field $\phi_{\pm}\propto e^{i\omega t}$ 
is given by
$\omega^2=k_3^2+2gB(n+1/2)\pm 2gB$.  
Integer $n\geq 0$ denote Landau levels 
and $k_3$ does
momentum parallel to the magnetic field.

The term $\pm 2gB$ in $\omega^2$ comes from the non-minimal interaction
which represents anomalous magnetic moment of the charged vector fields.
It is obvious that the modes of $\phi_{+}$ have imaginary frequencies $\omega^2=k_3^2-gB<0$ when $n=0$ and $k_3^2<gB$. 
The modes occupy the lowest Landau level ( $n=0$ ).
This implies that
the field $\phi_{+}$ increase exponentially in time.
The modes with $\omega^2<0$ are called as Nielsen-Olesen unstable modes.
Therefore,
when homogeneous color magnetic fields are present,
the states are unstable; the Nielsen-Olesen unstable modes are generated spontaneously
and the states decay into more stable states.
This is the Nielsen-Olesen instability in
the gauge theory.

For the convenience of later discussions, we write down the Hamiltonian\cite{iwazaki}
of the field $\phi_{+}$ neglecting higher order interactions,

\begin{equation}
\label{H}
H=|\partial_t\phi_{+}|^2+|\vec{D}\phi_{+}|^2-2gB|\phi_{+}|^2,
\end{equation}
where the term $|\vec{D}\phi_{+}|^2$ becomes $(k_3^2+gB)|\phi_{+}|^2$ 
when $\phi_{+}$ occupies the lowest Landau level under the homogeneous color magnetic field $B$.
The Hamiltonian holds even for inhomogeneous magnetic field $B=\epsilon_{i,j}\partial_iA_j$.
We can see that even when inhomogeneous magnetic field $B$ is present,
the presence of the field $\phi_{+}$
can make lower the energy than the energy ( $=0$ )of the state with $\phi_{+}=0$.
In the sence, inhomogeneous magnetic fields are also unstable in general. 
( In order to demonstrate the instability we need to show 
the presence of bound state solutions by solving eq(\ref{4}) in the next section. )
We call the field $\phi_{+}$ as Nielsen-Olesen field.
Similarly, the field $\phi_{-}$ is a Nielsen-Olesen field for the
inhomogeneous $B$ since it also describes unstable modes;
the Hamiltonian of $\phi_{-}$ is given by
$H(\phi_{-})=|\partial_t\phi_{+}|^2+|\vec{D}\phi_{+}|^2+2gB|\phi_{+}|^2$
and the sign of $gB$ can be negative in the transverse space.

\section{Our assumption for analysis of instability in gauge fields}

First of all,
we explain assumptions used in our simple model for gauge field evolution in heavy ion collisions.
Here we use the Cartesian coordinate for the explanation.
Initial gauge fields independent of the rapidity used in our model are only longitudinal color electric and magnetic fields.
They are maximal Abelian components of gauge fields.
For example, 
$A_i^a=(0,0,A_i^3)$ in the SU(2) gauge theory. Then, non-Abelian interactions vanish and only
linear equations remain. 
It has been discussed\cite{itakura} that even if we make such a simplification of initial gauge fields in the glasma, 
we can reproduce quite well the numerical results\cite{5} in their evolution at least for $\tau>Q_s^{-1}$.
Thus, the approximate use of such Abelian initial gauge fields instead of the non-Abelian glasma is 
appropriate. This is the first our assumption.

Such Abelian gauge fields are inhomogeneous in transverse space similarly as
the glasma. 
Then, as we have discussed in eq(\ref{H}), the color magnetic fields $B$
are unstable energetically
owing to the presence of Nielsen-Olesen fields.
This instability represents instability of the glasma in our model.
It is described by the equation,

\begin{equation}
\label{4}
\omega^2\phi=(-D_T^2+k_3^2-2gB)\phi
\end{equation}
with $D_T^2=(\vec{\partial}_T+ig\vec{A}_T)^2$ and $B=\mbox{rot}A_T$,
where we assume that $\phi\propto \exp(i\omega t-ik_3x_3)$.
The equation can be derived from the Hamiltonian in eq(\ref{H}).
It is a "Schr$\ddot{o}$dinger equation" of charged particles under the magnetic field $B$ with 
an additional potential term $-2gB$. 
There are solutions of "bound states".
The binding energy is given by $-\omega^2>0$. Thus,
the growth rate of the unstable modes is given by the imaginary part of the frequency $\omega$.

The second our approximation is to
replace inhomogeneous $B$ in the operator,
$-\vec{D}_T^2-2gB$ in eq(\ref{4}) with
effective homogeneous one $\bar{B}$, keeping
the eigenvalue of the operator $-\vec{D}_T^2-2gB$ unchanged. 
We should note that the dependence of the longitudinal momentum $k_3$ in
the growth rate and the momentum distribution $\phi(k_3)$ do not change even if we make
the replacement as far as eigenvalues of the operator, $-D_T^2-2gB$,
are the same. 

Thus, by the replacement 
we can analyze what we are concerned with, that is,
the time evolutions of the unstable modes and 
the characteristic features, e.g. growth rates, associated with the evolutions.
But local behaviors of the fields in the transverse space are lost 
by the replacement.
The replacement is possible in principle, but is difficult in practice.
In our simple model, the value of $\bar{B}$ is determined by
comparing our results with the previous simulation\cite{ve}.

\section{Our simple model of Nielsen-Olesen instability in glasma}
	
Now we explain the detail of our simple mode for the glasma decay.
We use the proper time coordinate, $\tau=\sqrt{x_0^2-x_3^2}$ and 
rapidity, $\eta=\log(\frac{x_0+x_3}{x_0-x3})$, as a longitudinal coordinate.
The coordinate is convenient for the description of expanding "glasma" generated in the early stage
of heavy ion collisions. The collisions occur at $x_3=0$ ( $\eta=0$ ) and $x_0=0$ ( $\tau=0$ )
in extremely high energies. Thus,
the heavy ions are Lorentz contracted to have vanishing width in the longitudinal direction and only extended in the transverse directions 
with coordinates $x_i=(x_1,x_2)$.

We analyze SU(2) gauge fields, $\vec{A}_{\mu}$.
Corresponding gauge fields in the coordinates are given such as $\vec{A}_{\tau}$, $\vec{A}_{\eta}$ and $\vec{A}_i=(\vec{A}_1,\vec{A}_2)$.
Taking a gauge condition, $\vec{A}_{\tau}=0$, the Lagrangian of the fields is given by

\begin{equation}
\tau L=\tau \Biggl( \frac{1}{2\tau^2}(\partial_{\tau}\vec{A}_{\eta})^2+\frac{1}{2}(\partial_{\tau}\vec{A}_i)^2
-\frac{1}{2\tau^2}\vec{F}_{\eta,i}^2-\frac{1}{4}\vec{F}_{i,j}^2 \Biggr) ,
\end{equation}
with $\vec{F}_{\eta,i}^2=(\partial_{\eta}\vec{A}_i-\partial_i\vec{A}_{\eta}+g\vec{A}_{\eta}\times \vec{A}_i)^2$
and $\vec{F}_{i,j}^2=(\partial_i\vec{A}_j-\partial_j\vec{A}_i+g\vec{A}_i\times \vec{A}_j)^2$.

We define the complex fields, $\phi_{\eta}$, $\phi_i$, and real ones, $A_{\eta}$, $A_i$ by rearranging 
color components of the gauge fields,

\begin{equation}
\vec{A}_{\eta}=(A_{\eta}^1,A_{\eta}^2,A_{\eta}^3)=
(\frac{\phi_{\eta}+\phi_{\eta}^{\dagger}}{\sqrt{2}},\frac{\phi_{\eta}-\phi_{\eta}^{\dagger}}{i\sqrt{2}},A_{\eta})
\quad \mbox{and} \quad \vec{A}_i=(\frac{\phi_i+\phi_i^{\dagger}}{\sqrt{2}},\frac{\phi_i-\phi_i^{\dagger}}{i\sqrt{2}},A_i),
\end{equation}
and define the following "charged field" with spin parallel, $\phi_{+}$ ( anti-parallel, $\phi_{-}$, ) to the longitudinal direction,
$\phi_{\pm}\equiv\frac{\phi_1\pm i\phi_2}{\sqrt{2}}$.
Initial gauge fields are introduced as maximal Abelian component $A_{\eta}, A_i$ of the gauge fields as we mentioned above.
It is easy to see that the fields, $\phi_{\eta}$ and $\phi_{\pm}$, are transformed under the gauge transformation,
$A_{\mu}\to U^{\dagger}A_{\mu}U+g^{-1}U^{\dagger}\partial_{\mu}U$ with $U=\exp(i\theta\sigma_3)$, such that
$\phi_{\eta,\pm}\to \exp( -i\theta)\phi_{\eta,\pm} $, where $\sigma_i$ are Pauli metrices and $\theta$ is a constant 
independent of $\tau$.
Therefore, we may think that the complex fields, $\phi_{\eta,\pm}$ are U(1) charged fields corresponding to the symmetry.

We introduce a longitudinal color magnetic field $B$ and an color electric field $E$ as initial background gauge fields, both of which are
assumed to point to the third direction in SU(2) gauge group; $B=\epsilon_{i,j}\partial_iA_j$ and $E=\frac{1}{\tau}\partial_{\tau}A_{\eta}$.
The fields point to the direction parallel to the collision axis in the real space; $\vec{B}=(0,0,B)$ and $\vec{E}=(0,0,E)$.
We assume that the background fields are generated at $\tau=0$ in heavy ion collisions.

With the use of the fields, $A_{i,\eta}$ and $\phi_{\pm,\eta}$, the Lagangian leads to

\begin{eqnarray}
\label{eq1}
\tau L&=&\frac{\tau}{2}(\partial_{\tau}A_i)^2+\frac{1}{2\tau}(\partial_{\tau}A_{\eta})^2+\tau(|\partial_{\tau}\phi_{+}|^2+|\partial_{\tau}\phi_{-}|^2)
+\frac{1}{\tau}|\partial_{\tau}\phi_{\eta}|^2-\frac{\tau}{4}f_{i,j}^2-\frac{1}{2\tau}f_{\eta,i}^2 \nonumber \\
&-&\tau(|D_i\phi_{+}|^2+|D_i\phi_{-}|^2)-\frac{1}{\tau}(|D_{\eta}\phi_{+}|^2+|D_{\eta}\phi_{-}|^2+|D_i\phi_{\eta}|^2)+2\tau gB(|\phi_{+}|^2-|\phi_{-}|^2) \nonumber \\
&+&\frac{\tau}{2}|D_{-}\phi_{+}+D_{+}\phi_{-}|^2+\frac{1}{\sqrt{2}\tau}((D_{-}\phi_{+}+D_{+}\phi_{-})D_{\eta}\phi_{\eta}+c.c.)
+\nonumber \\
&&\biggl(\frac{2gi}{\sqrt{2}\tau}(f_{\eta}\phi_{+}+f_{\eta}^{\dagger}\phi_{-})\phi_{\eta}^{\dagger}+c.c.\biggr)- 
\frac{g^2}{\tau}|\phi_{\eta}\phi_{+}^{\dagger}-\phi_{\eta}^{\dagger}\phi_{-}|^2-\frac{\tau g^2}{2}(|\phi_{+}|^2-|\phi_{-}|^2)^2,
\end{eqnarray}
with $f_{i,j}\equiv\partial_iA_j-\partial_jA_i$, $f_{\eta,i}\equiv\partial_{\eta}A_i-\partial_iA_{\eta}$, $f_{\eta}\equiv f_{\eta,1}-if_{\eta,2}$,
$D_i\equiv\partial_i+igA_i$, $D_{\eta}\equiv\partial_{\eta}+igA_{\eta}$, 
and $D_{\pm}\equiv D_1\pm iD_2$, where we have neglected surface terms just like $\partial_iJ_i$.

Obviously, the Lagrangian is invariant under the U(1) gauge transformation, $\phi_{\pm,\eta}\to \phi_{\pm,\eta}\exp(-i\theta)$
along with $A_{i,\eta}\to A_{i,\eta}+g^{-1}\partial_{i,\eta}\theta$. 
The kinetic energies of the fields $A_i$, $A_{\eta}$, $\phi_{\pm}$ and $\phi_{\eta}$,
are presented in the first line of the Lagrangian. 
In the second line, minimal interactions between $\phi_{\pm,\eta}$ and the abelian gauge fields,
$A_i$ and $A_{\eta}$ are presented.
We can see in the second line that the charged fields $\phi_{\pm}$ receive anomalous magnetic moments, $2\tau gB(|\phi_{+}|^2-|\phi_{-}|^2)$.
This term plays an important role for making the field $\phi_{+}$ unstable, that is, Nielsen-Olesen unstable mode.
There are the quartic interactions of the fields in the fourth line, 
which describe repulsive forces among the fields $\phi_{\eta,\pm}$.
The repulsive force leads to the saturation of the exponential increase observed in the simulation\cite{ve}.
The terms in the third line are irrelevant to our discussion below ( the terms can be gauged away ).

When an initial gauge field configuration of $B=\epsilon_{i,j}\partial_iA_j$ and $E=\frac{1}{\tau}\partial_{\tau}A_{\eta}$ is given at $\tau=0$,  
the subsequent evolution of the fields $A_{\eta}$ and $A_i$
is governed by the equations,

\begin{eqnarray}
&&\partial_{\tau}(\frac{1}{\tau}\partial_{\tau}A_{\eta})-\frac{1}{\tau}(\partial_i^2A_{\eta}-\partial_{\eta}\partial_iA_i)=0, \nonumber \\
\mbox{and} \quad &&\partial_{\tau}(\tau\partial_{\tau}A_i)-\tau(\partial_j^2A_i-\partial_i\partial_jA_j)
+(\partial_{\eta}^2A_i-\partial_i\partial_{\eta}A_{\eta})=0.
\end{eqnarray}
The equations are obtained from the Lagrangian by neglecting the other charged fields.
The approximation is valid when the other fields are sufficiently small so that the interactions between
$E$, $B$ and the others can be neglected.
A solution of $B$ and $A_{\eta}$ independent of the rapidity $\eta$ is given by
$B=B_0J_0(Q_0\tau)\cos(\vec{Q}_0\vec{x})$ and $A_{\eta}=c\tau J_1(Q_0\tau)\cos(\vec{Q}_0\vec{x})$,
where we suppose that the fields carry a transverse momentum, $\vec{Q}_0$ ( $Q_0=|\vec{Q}_0|$ )
as a typical transverse momentum of backgound gauge fields. 
$B_0$ is a constant and $J_{0,1}(Q_0\tau)$ are Bessel functions. 
( General solutions are given by the average in $Q_0$ over momentum distritutions. )
A constant of $c$ may be determined by the
requirement that $B=E$ as $\tau\to 0$. It leads to $c=B_0/Q_0$. The requirement arises from the initial condition of 
$\langle\mbox{Tr}(B^2)\rangle=\langle\mbox{Tr}(E^2)\rangle$ 
for $\tau\to 0$ given in the MV model of CGC \cite{fu}. 
Here, the expectation of $\langle\sim \rangle $ is taken over the distribution of
large x gluons according to the MV model.

Here we neglect spatial dependence of the gauge fields according to the second our
assumption. Additionally we simplify the factor $J_0(Q_0\tau)$
such that $J_0(Q_0\tau)\to\sin(Q_0\tau)\sqrt{2/\pi Q_0\tau}\sim \sqrt{2/\pi Q_0\tau}$ by neglecting the oscillating factor
$\sin(Q_0\tau)$. This smooth decay of the fields roughly coincides with
the numerical evolutions of the glasma\cite{5}. Hence,  
the simplification is appropriate for discussing small fluctuations
around the slowly decaying background gauge fields.
Therefore, we assume the following background initial gauge fields,

\begin{equation}
\label{sim}
B=B_0\sqrt{2/\pi Q_0\tau}  \quad \mbox{and} \quad 
A_{\eta}=B_0(\tau/Q_0)\,\sqrt{2/\pi Q_0\tau},
\end{equation}
which reproduce the smooth decay of $\langle\mbox{Tr}(B^2)\rangle$ and $\langle\mbox{Tr}(E^2)\rangle$
for large $\tau$.

Under the background gauge fields, 
we analyze the development of the small fluctuations $\phi_{\eta,\pm}$.
The fluctuations correspond to the small fluctuations added to initial background gauge fields
in the previous simulation\cite{ve}.
Since they are supposed to be much small, 
we take into account only quadratic terms of the fields in the Lagrangian.
These fluctuations in general oscillate with small amplitudes.
But one of these fluctuations increases exponentially. 
In our model the field $\phi_{+}$ ( or $\phi_{-}$  when $gB<0$ ) is the one increasing exponentially with $\tau$. The other fields 
simply oscillate and their amplitudes remain small.
	


We write down the equation of motion of the field $\phi_{+}$,

\begin{equation}
\label{5}
\partial_{\tau}^2\phi_{+}+\frac{1}{\tau}\partial_{\tau}\phi_{+}+
\biggl(\frac{(k_{\eta}-gA_{\eta}(\tau))^2}{\tau^2}-gB(\tau) \biggr)\phi_{+}=0,
\end{equation}
where $k_{\eta}$ denotes a longitudinal momentum; $\phi_{+}\propto\exp(-ik_{\eta}\eta)$.
We have taken 
only a component in the lowest Landau level. 
It is easy to see that
due to the last term $-gB(\tau)$, $\phi_{+}$ increases exponentially just as $\phi_{+}\propto\exp(\sqrt{gB}\tau)$ for $\tau \to \infty$
when $gB$ is independent of $\tau$, i.e. a solution of the equation, $\partial_{\tau}^2\phi_{+}-gB\phi_{+}\simeq 0$.
Their wave functions are given such that $\phi_{+}=g_m(\tau) z^m\exp(-|\vec{x}|^2/4l^2_B) $ with $z=x_1+ix_2$ and integers, $m\ge 0$,
where $l_B=1/\sqrt{gB}$ denotes cyclotron radius and $g_m(\tau)$ is governed by the eq(\ref{5});
we have neglected the smooth expansion of the cyclotron radius.

\section{Our results}
In order to solve the eq(\ref{5}) with $A_{\eta}(\tau)$ and $B(\tau)$ given in eq(\ref{sim}), 
we rewrite the equation in the following,

\begin{equation}
\label{8}
\partial_{\tau'}^2\phi_{+}+\frac{1}{\tau'}\partial_{\tau'}\phi_{+}+
\biggl(\frac{(k_{\eta}-b\sqrt{\tau'})^2}{\tau'^2}-\frac{a}{\sqrt{\tau'}}\biggr)\phi_{+}=0,
\end{equation}
where dimensionless parameters are defined as $\tau'\equiv Q_s\tau$, $a\equiv \sqrt{2/\pi}(gB_0/Q_0^2)\times (Q_0/Q_s)^{3/2}$ 
and $b\equiv \sqrt{2/\pi}(gB_0/Q_0^2)\times (Q_0/Q_s)^{1/2}$.
In the subsequent calculations we treat the scale of the field $\phi_{+}$ arbitrary 
although it is much smaller
than the background field. This is allowed in the approximation of
taking only quadratic terms of the field $\phi_{+}$ in the Lagrangian.
The coefficients $a$ and $b$ are determined for reproducing the results in the previous
simulation\cite{ve}; actually we have used $a=(0.05)^2$ and $b=0.38$
in order to obtain our curve in Fig.5.

Before solving the equation (\ref{8}) numerically, we briefly explain how the solutions behave with $\tau'$.	
The term of $\biggl(\frac{(k_{\eta}-b\sqrt{\tau'})^2}{\tau'^2}-\frac{a}{\sqrt{\tau'}}\biggr)$ ( $\equiv \omega_s^2$ ) is 
just a spring constant. The term of $\frac{1}{\tau'}\partial_{\tau'}\phi_{+}$ represents a friction, which
becomes weaker as $\tau'$ becomes larger.
Thus, we understand that the field oscillates as far as $\omega_s^2 >0$, that is, in the early stage ( $\tau'\sim O(1)$ )
after the production of the background fields. The spring constant $\omega_s^2$ becomes small with $\tau'$. 
Once $\omega_s^2$ becomes negative, the field stops the oscillation
and begins to increase exponentially.


In Fig.1 we show the typical behavior of the field, $\phi_{+}(k_{\eta}=16.5,\tau')$, with 
the initial conditions of $\phi_{+}(\tau'=0.01)=1$ and $\partial_{\tau}\phi_{+}(\tau'=0.01)=0$.
( Obviously, taking different initial conditions do not change the global behavior of $\phi_{+}$ for large $\tau'$,
since it simply oscillates in small $\tau'$. )  
The field increases exponentially after the oscillation in the early stage. 
When the longitudinal momentum $k_{\eta}$ becomes large, 
the time $\tau$ when the field $\phi_{+}(k_{\eta},\tau')$
begins to increase exponentially,
becomes large. This implies that as $\tau'$ becomes larger, 
the modes with larger longitudinal momentums
are excited. 

In Fig.2 we show the time dependence of the maximal momentum $k_{\eta}(\mbox{max})$. 
The maximal momentum $k_{\eta}(\mbox{max})$ at the time $\tau'$ is defined as the momentum with which
the mode $\phi_{+}(k_{\eta}(\mbox{max}),\tau')$ starts to increase 
exponentially at the time $\tau'$. The modes with $k_{\eta}<k_{\eta}(\mbox{max})$ 
have already been increasing exponentially at the time $\tau'$.
The maximal momentum $k_{\eta}(\mbox{max})$ may be obtained by solving the condition of
the spring constant $\omega_s^2=0$, but 
$k_{\eta}(\mbox{max})$ is defined as $|\phi_{+}(k_{\eta}(\mbox{max}),\tau')|=2$ in our calculation.
We have shown both results in Fig.2, which almost coinside with each other.
$k_{\eta}(\mbox{max})$ increases almost linearly in $\tau'$, but
the solution of $\omega_s^2=0$ shows that $k_{\eta}(\mbox{max})\propto \tau'\,^{3/4}$. 
The result agrees with the previous one\cite{ve}, 
although the rate of the increase is approximately four times smaller than the previous one;
$k_{\eta}(\mbox{max})\simeq 0.015\,Q_s\tau+5$. 

It has been shown\cite{ve} that
$k_{\eta}(\mbox{max})$ deviates from the linear dependence around the time when
the exponential increase is saturated. After the deviation, $k_{\eta}(\mbox{max})$ increases 
very rapidly. These phenomena could be understood in our model
as the result due to the onset of quartic interactions among the field $\phi_{+}$.
The quartic interactions make the energy of the mode with largest amplitude
be transmitted to the other modes with higher longitudinal momenta.

\begin{figure}[htb]
\begin{minipage}{.47\textwidth}
\includegraphics[width=6cm,clip]{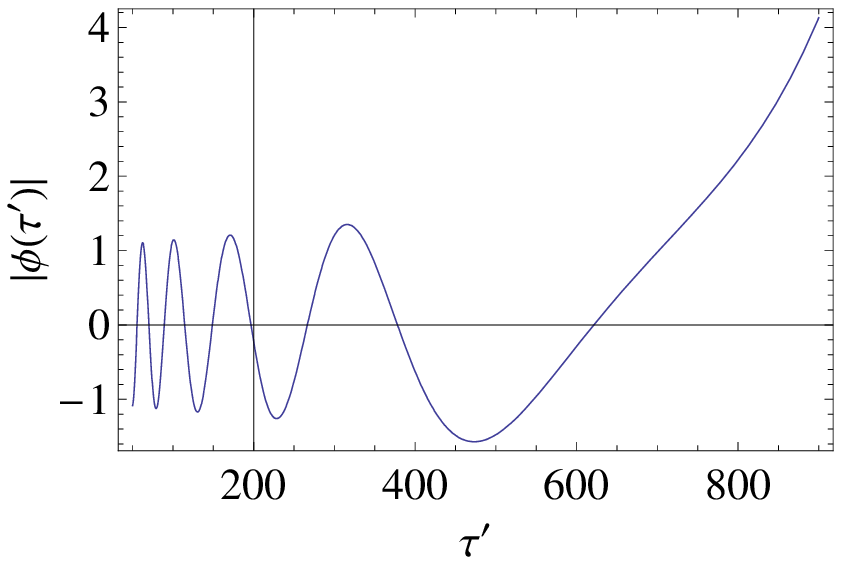}
\label{fig:t-dependence}
\caption{After the oscillation, $\phi_{+}(k_{\eta}=16.5)$ increases exponentially with $\tau'=Q_s\tau$.}
\end{minipage}
\hfill
\begin{minipage}{.47\textwidth}
\includegraphics[width=6cm,clip]{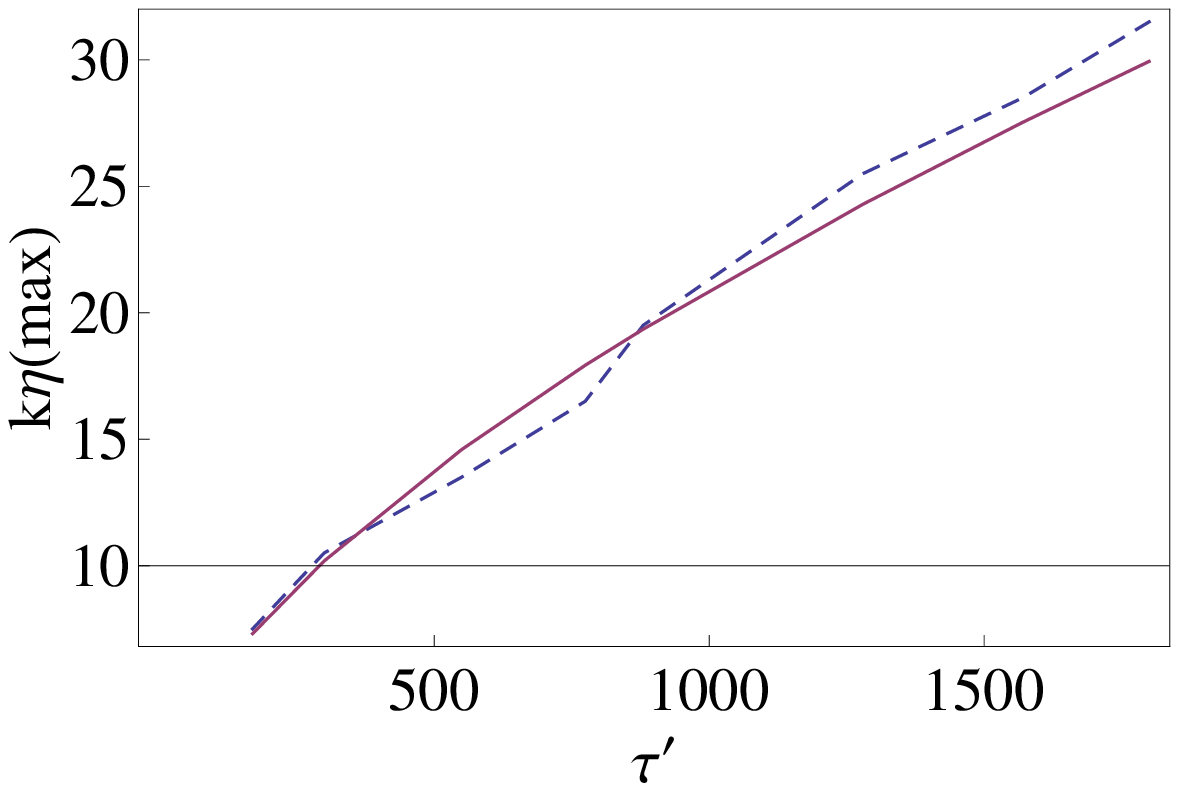}
\caption{solution of $\omega_s^2=0$ ( solid ),
$k_{\eta}(\mbox{max})$ ( dashing ) increases almost linearly with $\tau'=Q_s\tau$.}
\label{fig:kmax}
\end{minipage}
\end{figure}

Before proceed to make the comparison, we should mention 
why we compare the evolution of the field $\phi_{+}$ with 
the evolution of longitudinal pressure discussed in the previous simulation\cite{ve}.
Their behaviors are quite similar to each other.
Roughly speaking, there are two components of 
fields involved in the ref.\cite{ve}, large ones and much small ones. 
The large ones $A(\mbox{large})$ are just the background
fields independent of rapidity, which are produced according to 
the MV model.
On the other hand, the small ones $a(\mbox{small})$ are fluctuations depending on the rapidity 
which are added by hand to the background fields.
In the circumstance,
the longitudinal pressure $P_{\eta}$    
is composed of two parts, $P_{\eta,0}$ and $\delta P_{\eta}$; $P_{\eta}=P_{\eta,0}+\delta P_{\eta}$.

\begin{equation}
P_{\eta}=\tau^{-2}\biggl(\vec{F}_{\eta,i}^2+(\tau\partial_{\tau}\vec{A}_i)^2\biggr)-
\vec{F}_{1,2}^2-(\frac{1}{\tau}\partial_{\tau}\vec{A}_{\eta})^2\simeq P_{\eta,0}+\delta P_{\eta},
\end{equation}
where $P_{\eta,0}$ is rapidity independent
and is formed only of the large components, while $\delta P_{\eta}$ is formed of 
the small components as well as the large ones;
$\delta P_{\eta}=P(A(\mbox{large}))\times a(\mbox{small})$.
The linear dependence on $a(\mbox{small})$ comes from
the approximation
of $a(\mbox{small}) \ll A(\mbox{large})$.
$P_{\eta,0}$ decreases smoothly with $\tau$.
On the other hand, $\delta P_{\eta}$  
increases exponentially although it is still
much smaller than $P_{\eta,0}$.
Therefore, some of small components $a(\mbox{small})$ depending on the rapidity,
increase exponentially as has been shown\cite{ve}. Our simple model of gauge field evolution indicates that such a small component
is just the Nielsen-Olesen unstable mode $\phi_{+}$. 
That is the reason why we compare the evolution of the field $\phi_{+}$ with
the evolution of the longitudinal pressure. 
It should be noted that the Fourier component of $\delta P_{\eta}$ in the rapidity
is determined only by the factor of $a(\mbox{small})$, which corresponds to small fluctuations $\phi_{\eta,\pm}$
in our model.

We now proceed to make the further comparison.
In Fig.3, we show the typical momentum distribution $|\phi_{+}(k_{\eta},\tau')|$ at $\tau'=1500$.
The distribution in $k_{\eta}$ has been obtained by 
solving eq(\ref{8}) for $\phi_+(k_{\eta},\tau')$ with each $k_{\eta}$ chosen within the range
 $0.24\le k_{\eta}\le 18$ 
by $\delta k_{\eta}=0.03$. 
The distribution is not smooth but oscillating rapidly in $k_{\eta}$. 
( When we magnify a small region, e.g. $\delta k_{\eta}\sim 0.2$ of the Fig.3, 
we can see explicitly the oscillation in $k_{\eta}$. Roughly speaking, this oscillation comes from 
the oscillation of a modified Bessel function
$I_{k_{\eta}}(Q_s\tau)$. )
But we can 
find out a smooth distribution obtained by averaging the original one over 
an appropriately small but sufficiently large $\delta k_{\eta}$
to be compared with wave length in the oscillation. Then, the smooth one coincides almost with 
the mountain-like form of the distribution in Fig.3.
( The average corresponds to an average over initial conditions for solving
eq(\ref{8}). This is because slightly different initial conditions lead to slightly different
curves. )
We find that the distribution has a peak at a momentum $k_{\eta}(\rm{p})$.
The mode with the momentum is generated most efficiently. 

In Fig.4, we show how $k_{\eta}(\rm{p})$
depends on the time $\tau'$; 
$k_{\eta}(\rm{p})$ increases very slowly with $\tau'$ just as $0.18 \sqrt{\tau'}$.
Furthermore, we can see that the momenta
are smaller than $k_{\eta}(\mbox{max})$. These results agree with
the previous ones\cite{ve}.
The presence of the longitudinal momentum $k_{\eta}(\rm{p})\simeq 6\sim 8$
almost independent of time implies that
in the decay of the gauge fields uniform in $\eta$,
specific modes with the momentum $k_{\eta}(\rm{p})$ is generated most efficiently,
which breaks the homogeneity in $\eta$.
We do not understand why such specific momenta are present.

\noindent

\begin{figure}[t]
\begin{minipage}{.47\textwidth}
\includegraphics[width=6cm,clip]{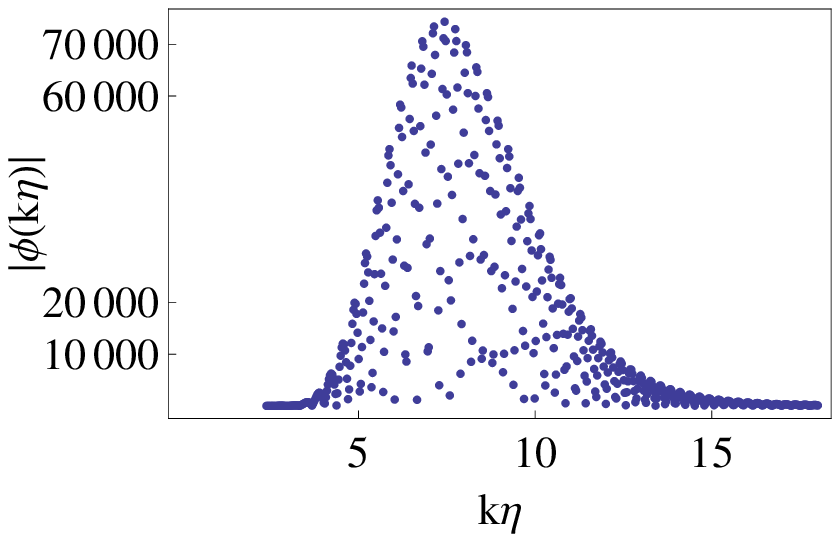}
\label{fig:distribution}
\caption{distribution of longitudinal momentum $k_{\eta}$ at $\tau'=Q_s\tau=1500$.}
\end{minipage}
\hfill
\begin{minipage}{.47\textwidth}
\includegraphics[width=6cm,clip]{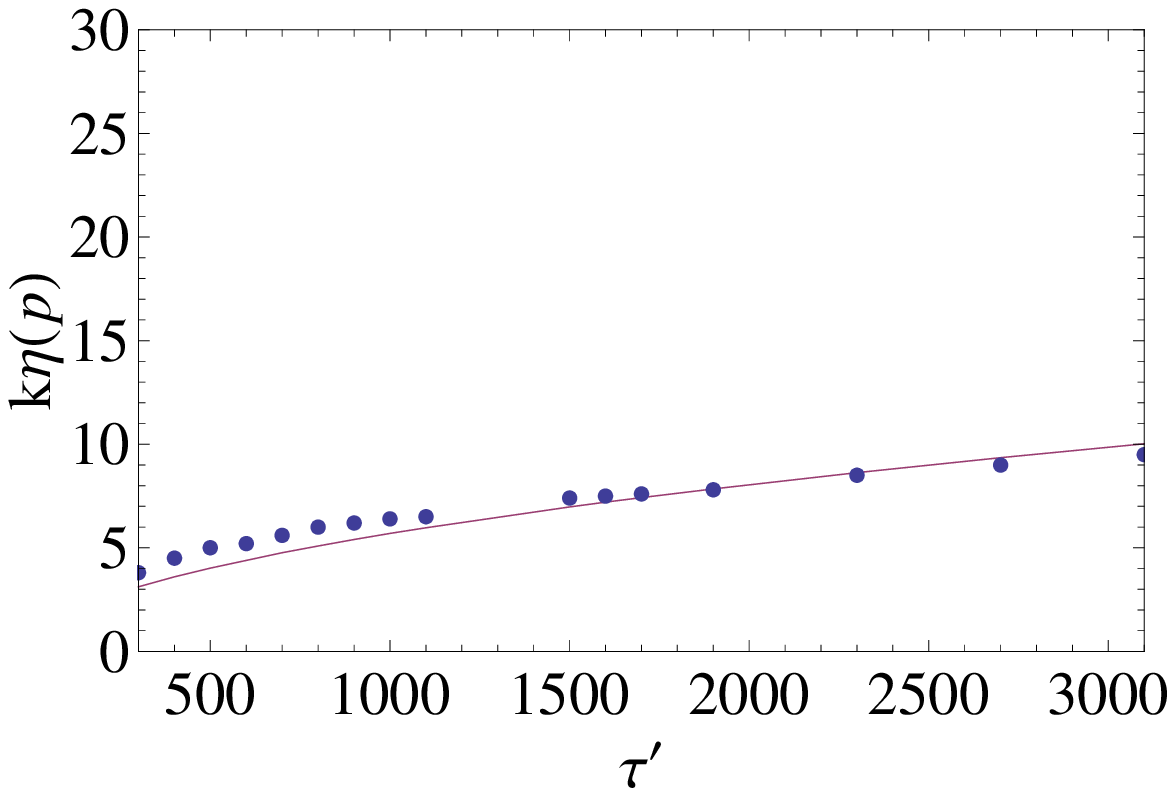}
\caption{ $\tau'$ dependence of $k_{\eta}(p)$ ( dot ) and 
$k_{\eta}=0.18\sqrt{\tau'}$ ( solid ).}
\label{fig:k(p)}
\end{minipage}
\end{figure}


Finally, we show in Fig.5 how $|\phi_{+}(k_{\eta}(\rm{p}),\tau)|$ increases
with $\tau'$, that is, the time dependence of the peak $|\phi_{+}(k_{\eta}(\rm{p}),\tau)|$. 
It increases as $\exp(\tau'\,^{3/4})$ for 
$\tau'\to \infty $, 
which can be read
from the equation(\ref{8}). The growth rate is defined as $(\log|\phi|)/\tau$.
Thus, the growth rate of the field $|\phi_{+}(k_{\eta}(\rm{p}),\tau)|$ 
decreases slowly with $\tau$.
Since $k_{\eta}(\rm{p})$ is almost constant in time, 
$|\phi_{+}(k_{\eta},\tau)|$ also increases
in the similar way to $|\pi_{+}(k_{\eta}(\rm{p}),\tau)|$ for any $k_{\eta}$.

For a comparison, we have depicted the Fourier component of the longitudinal pressure, 
$P_{\eta}(k_{\eta}(p))=(d_0+d_1\exp(0.427\sqrt{\tau'}))$ in the ref.\cite{ve}
as well as the function of $\exp(0.00544\tau')$ also used in the ref.\cite{ve}. 
Obviously,
the field, $|\phi_{+}(k_{\eta}(\rm{p}),\tau')|$, in our calculation agrees with 
the longitudinal pressure
better than 
the function of $\exp(0.00544\tau')$. 
( In Fig.5, we take the scale arbitrary in the vertical coordinate. 
In order for $|\phi_{+}(k_{\eta}(\rm{p}),\tau')|$ to coincide precisely with the 
pressure shown in the ref.\cite{ve}, we have only to take the scale of the field
$|\phi_{+}(k_{\eta}(\rm{p}),\tau')|$ appropriately. ) 
Our simulation does not reproduce strictly the behavior of the pressure such as $\exp(\sqrt{\tau'})$. 
We expect that the more elaborate treatment of the background magnetic field may
give rise to the behavior $\exp(\sqrt{\tau'})$. ( When $B$ decreases as $\sim 1/\tau'$ instead of $1/\sqrt{\tau'}$,
this behavior can be obtained. We will discuss 
the validity of the behavior $B\sim 1/\tau'$ in near future. )

\noindent

\begin{figure}[t]
\begin{minipage}{.47\textwidth}
\includegraphics[width=6cm,clip]{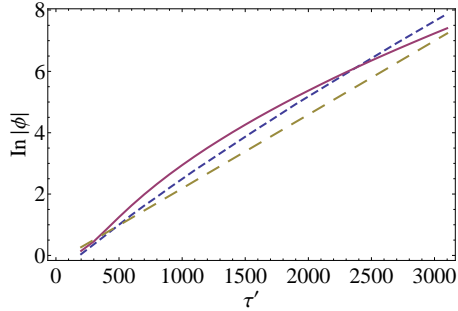}
\label{fig:comparison2}
\caption{ longitudinal pressure $P_{\eta}(k_{\eta}(p))$ ( solid ) in ref.\cite{ve} , $|\phi_{+}|$ ( short dashing )
and $\exp(0.00544\tau')$ ( dashing ).}
\end{minipage}
\end{figure}


As it has been shown, the behaviors of the longitudinal
pressure calculated in the MV model with the small
fluctuations added can be roughly reproduced in our simple model:
1) $k_{\eta}(\mbox{max})$ increases linearly with $\tau$, 2) $k_{\eta}(\rm{p})$
depends on $\tau$ very weakly and $k_{\eta}(\rm{p})$ is much smaller than $k_{\eta}(\mbox{max})$, 
and 3) the pressure $P_{\eta}(k(p))$ 
increases with $\tau$ as $\exp(\tau^{3/4})$, although the pressure obtained in the simulation 
increase as $\exp(\sqrt{\tau})$.
Furthermore, we may argue that the saturation of the exponential increase arises due to
the repulsive self-interaction of $\phi_{+}$ in our model. 
The repulsive interaction need more energies for the field to increase more.
Thus, it would stop the field increasing. 
Although our results are different in detail with those in the simulation, 
the rough agreement shows that 
our simple model of instabilities
is valid approximation for the instabilities 
observed in the glasma.
Thus, 
the instabilities of the glasma observed in the
previous simulation is caused by
the Nielsen-Olesen unstable mode.

In order to obtain these results,
we have used the parameters such as $a=(0.05)^2$ and $b=0.38$.
Roughly speaking, the parameter $a$ gives the growth rate of the
field $\phi_+$. Thus, the parameter can be determined by making it fit the growth rate 
obtained in the simulation\cite{ve}. But the growth rate shown in the simulation
is the one of the field $\phi_+(k_{\eta}(p))$, not $\phi_+(k_{\eta})$ itself. 
Here $k_{\eta}(p)$ depends on
$\tau$ although the dependence is very weak. This requires 
a careful adjustment of the other parameter $b$. 

The determination of the parameters in detail is not
important. An important thing is that these parameters lead to
a weak homogeneous color magnetic field $B_0$.
Actually,
the above values of $a$ and $b$ corresponds to the physical parameters of $Q_0\simeq Q_s/152$ and $gB_0\simeq 4.7Q_0^2$.
Thus, the growth rate $\sim \sqrt{gB_0}$ discussed in the section (4)
is much smaller than $\sqrt{g|B|}\sim Q_s$. Indeed, the growth rate is about
$0.00544Q_s$ derived from the reference function $\exp(0.00544\tau')$ depicted in Fig.5.

We should mention that the smallness of the growth rate comes from the inhomogeneity of the glasma.
The potential $-2gB\sim Q_s^2$ for the field $\phi_{+}$ in eq(\ref{4})
varies so rapidly in transverse space; it has many attractive regions ( $-2gB<0$ ) and
repulsive regions ( $-2gB>0$ ) whose widths are of the order $Q_s^{-1}$. 
Wave functions of the bound states are extended over these regions
involving both attractive and repulsive potentials; they can never be trapped
within a region with attractive potential.
Therefore, the binding energies $-\omega^2$
become much smaller than $Q_s^2$.
Thus, the growth rate becomes much smaller than $Q_s$.

\section{Conclusion}
To summarize,
we have discussed that inhomogeneous color magnetic fields $gB\sim Q_s^2$ produced in the 
high energy heavy ion collisions decay with the production of 
the Nielsen-Olesen fields $\phi_{+}$. 
Instead of analyzing the evolution of the field $\phi_{+}$ under the inhomogeneous
color magnetic fields, we analyze it
using an effective homogeneous color magnetic field.
Then, 
we have compared the time evolution of
the Nielsen-Olesen field $\phi_{+}$
with the evolution of
the longitudinal pressure shown in the simulation\cite{ve}. 
We have found that our simple model with the effective weak homogeneous color magnetic field $gB_0$
reproduces the important features clarified in the simulation;
growth rates, longitudinal momentum distributions, etc. of the unstable modes.   
The coincidence is not accidental.
These are properties associated with time and longitudinal directions.
Even if we use the homogeneous magnetic field, the properties of the unstable modes can be
reproduced in principle. On the other hand, 
properties associated with the transverse directions
can not be reproduced with the use of such fields.   
Therefore, our analysis shows that the decay of the glasma
generated initially at $\tau=0$ in heavy ion collisions
is caused by Nielsen-Olesen instability.

On the other hand, there is an insistence\cite{ve,arnold} that
Weibel instability known in plasma physics
is the cause of the glasma instability:
Inhomogeneous electromagnetic plasma whose momentum distribution depends only on transverse momentum,
shows Weibel instability under small magnetic field applied.  
The instability is discussed by using
the Boltzmann equation of charged ( color charged ) particles 
coupled with electromagnetic ( color gauge ) fields,
while the glasma instability in the simulation has been
shown in pure gauge theory.
Nielsen-Olesen instability is the instability
arising in pure gauge theory. In this sense, the relevance of the Weibel instability
to the glamsa instability is not obvious.
Thus, it is reasonable to think
that the glasma instability shown in the simulation
is just Nielsen-Olesen instability. 
We will discuss in future publications
why the glamsa instability is just Nielsen-Olesen instability,
not Weibel one.

\vspace*{2em}
We would like to express thanks
to Dr.K. Itakura in KEK and Dr.H. Fujii in University of Tokyo for useful comments.


%


\begin{thebibliography}{99}
\bibitem{1}L.D. McLerran and R. Venugopalan, Phys. Rev. D49, 2233 (1994);
D49, 3352 (1994); D50, 2225 (1994).
\bibitem{2}A. Krasnitz and R. Venugopalan, Nucl. Phys. B557, 237 (1999);
Phys. Rev. Lett. 84, 4309 (2000); 86, 1717 (2001).
\bibitem{3}A. Krasnitz, Y. Nara and R. Venugopalan, Phys. Rev. Lett. 87, 192302 (2001);
Nucl. Phys. A717, 268 (2003).
\bibitem{4}T. Lappi, Phys. Rev. C67, 054903 (2003); C70, 054905.
\bibitem{5}T. Lappi and L. McLerran, Nucl. Phys. A772, 200 (2006).
\bibitem{cgc}E. Iancu, A. Leonidov and L. McLerran, hep-ph/0202270.\\
E. Iancu and R. Venugopalan, hep-ph/0303204.
\bibitem{hirano}T. Hirano and Y. Nara, Nucl. Phys. A743, 305 (2004); J. Phys. G30, S1139 (2004).
\bibitem{iwa}A. Iwazaki, Phys. Rev. C77. 034907 (2008). 
\bibitem{ve}P. Romatschke and R. Venugopalan, Phys. Rev. Lett. 96, 062302 (2006); Phys. Rev. D74, 045011 (2006).
\bibitem{fuku}K. Fukushima, F. Gelis and L. McLerran, Nucl. Phys. A786, 107 (2006).
\bibitem{review}F. Gelis, T. Lappi and R. Venugopalan, hep-ph/0708.0047.
\bibitem{no}N.K. Nielsen and P. Olesen, Nucl. Phys. B 144 (1978) 376;
Phys. Lett. B 79 (1978) 304.
\bibitem{savvidy}G.K. Savvidy, Phys. Lett. B 71 (1977) 133.\\
H. Pagels, Lecture at Coral Gables, Florida, 1978.
\bibitem{itakura}Y.V. Kovchegov, Nucl. Phys. A762 298 (2005).\\
T. Lappi, Phys. Lett. B643 11 (2006).\\
H. Fujii and K. Itakura, Nucl. Phys. A809 88 (2008).\\
H. Fujii, K. Fukushima and Y. Hidaka, hep-ph/0811.0437.
\bibitem{miklos}M. Gyulassy and A. Iwazaki, Phys. Lett. 165B 157 (1985).\\
N. Tanji, hep-ph/08104429, see references in the paper.
\bibitem{abelian}Z.F. Ezawa and A. Iwazaki, Phys. Rev. D25 2681 (1982).
\bibitem{iwazaki}A. Iwazaki, Phys. Rev. D75:034020 (2007).
\bibitem{fu}K. Fukushima, Phys. Rev. C76, 021902(R) (2007).
\bibitem{arnold}P. Arnold and G.D. Moore, Phys.Rev. D76:045009 (2007).
\end{thebibliography}
\end{document}